\documentclass[letterpaper, 10 pt, conference]{ieeeconf}  

\IEEEoverridecommandlockouts                              
\usepackage[utf8]{inputenc}
\usepackage[T1]{fontenc}
\usepackage{hyperref}

\title{\LARGE \bf
VOTE400(Voide Of The Elderly 400 Hours): A Speech Dataset to Study Voice Interface for Elderly-Care
}

\author{Minsu Jang$^{1}$, Sangwon Seo$^{2}$, Dohyung Kim$^{1}$, Jaeyeon Lee$^{1}$ and Jaehong Kim$^{1}$ and Jun-Hwan Ahn$^{2}$
\thanks{$^{1}$Minsu Jang, Dohyung Kim, Jaeyeon Lee and Jaehong Kim are with Electronics and Telecommunications Research Institute, Daejeon-si, South Korea
        {\tt\small minsu at etri.re.kr}}%
\thanks{$^{2}$Sangwon Seo and Jun-Hwan Ahn is with MINDs Lab Inc., Kyungki-do, South Korea
        {\tt\small asdn9353 at mindslab.ai}}%
}

\begin{document}

\maketitle
\thispagestyle{empty}
\pagestyle{empty}

\begin{abstract}

This paper introduces a large-scale Korean speech dataset, called VOTE400, that can be used for analyzing and recognizing voices of the elderly people. The dataset includes about 300 hours of continuous dialog speech and 100 hours of read speech, both recorded by the elderly people aged 65 years or over. A preliminary experiment showed that speech recognition system trained with VOTE400 can outperform conventional systems in speech recognition of elderly people's voice. This work is a multi-organizational effort led by ETRI and MINDs Lab Inc. for the purpose of advancing the speech recognition performance of the elderly-care robots.

\end{abstract}

\section{INTRODUCTION}

Voice interface is the most intuitive, comfortable and universal interface for interacting with service robots. Recent advancement of commercial cloud-based speech-to-text (STT) services allowed devising a voice interface for service robots a very simple process of integrating a service API for STT into the robot SW system.

While these commercial systems work very well with adults in the ages of between 20 and 60, it easily fails with voices from older adults aged 65 years or over. It is known that speech signals from older adults bring about difficulties for automated speech recognition as they tend to be imprecise in consonant pronunciation, include tremors, and have slower articulations \cite{c1}.

In the need to develop a speech recognition system that are specialized to the speech signals from older adults, we built a speech dataset by collecting large-scale dialogue and read speech from older adults. The result of our effort is 400 hours of Korean speech data which we named as 'VOTE400 (\textbf{V}oice \textbf{O}f \textbf{T}he \textbf{E}lderly \textbf{400} Hours) and open-sourced for any non-commercial research projects (https://ai4robot.github.io/mindslab-etri-vote400).

\section{Dataset Description}

\subsection{Dataset Collection}

For recruiting and collecting voice data from older adults, we got assistance from a Korean governmental office, called Dok-Geo-No-In-Jong-Haap-Ji-Won Center (Ji-Won Center: \url{http://www.1661-2129.or.kr/index.html}) devoted to the support of older adults living alone. With the support from the Ji-Won Center, we could collect a large-scale dialog speech and read speech from a number of older adults across various regions of South Korea.

\subsubsection{Dialog Speech}
To collect spontaneous speech data from older adults, we could utilize a support program of Ji-Won Center called Saa-Raang-It-Gi where social workers regularly visit elderly people's homes for consulting on health-related issues and relieving loneliness. After explaining about the data collection experiment and getting consent of participation from the elderly, conversations between a social worker and an elderly were recorded using a smartphone. 

The recordings from these program sessions were sent to Ji-Won Center and a screening process was performed to remove every dialogue involving sensitive personal information. Then, a quality assurance process was followed to filter out speech segments incomprehensible by human listener due to imprecise pronunciation or significant noise.

\subsubsection{Read Speech}
To amend the relatively low-quality of the dialog speech dataset, we launched another data collection process to acquire read speech from older adults. We built and utilized in the process a dedicated speech collection system, where a tablet-based client program presents a sentence to read to an elderly user; makes a recording and sends it to a server; where the recording is inspected to be accepted or not. In total, the number of unique sentences chosen to be read by participants was 2,250. These sentences were selected by considering how often these could be casually uttered by older adults in daily lives.

\begin{table}[h]
\caption{Raw Data Collection of Dialog Speech}
\label{regional_dist_participants}
\begin{center}
\begin{tabular}{lrr}
\hline
Region(\texttt{R})            & No. Participants    & Len. (hrs)  \\ \hline \hline
Seoul-si(SE)          & 620                 & 122                   \\ \hline
Busan-si(PS)          & 242                 & 90                   \\ \hline
Daegu-si(DG)      & 202                 & 33                   \\ \hline
Gwangju-si(GJ)      & 179                 & 63                   \\ \hline
Daejeon-si(DJ)        & 275                 & 66                   \\ \hline
Ulsan-si(WS)      & 80                  & 28                   \\ \hline
Goyang-si(GG)      & 335                 & 69                   \\ \hline
Gangwon-do(GW)      & 178                 & 45                   \\ \hline
Chungcheongbuk-do(CB) & 252                 & 92                   \\ \hline
Chungcheongnam-do(CN) & 317                 & 46                   \\ \hline
Jeollanam-do(JN)      & 323                 & 103                   \\ \hline
Gyeongsangbuk-do(GB)  & 378                 & 116                   \\ \hline \hline
Total             & 3,381               & 873                   \\ \hline
\end{tabular}
\end{center}
\end{table}

\subsection{Dialog Speech Data}

The total number of elderly participants is 3,381 and the total length of recordings is 873 hours. This is the result of collective efforts by regional senior citizens welfare institutes collaborating with the Ji-Won Center. Table \ref{regional_dist_participants} shows the regional distributions of all the participants and the length of recordings per region.

After the screening and the QA process mentioned in the previous subsection, we finalized 300 hours of dialog speech to be included in the VOTE400 dataset. Transcription for every final speech data was done by human annotators. In VOTE400, we provide for a recording session in a WAV file. The audio format of the WAV file is as shown in table \ref{dialog_speech_audio_format}. Every WAV file is accompanied by a transcription text file encoded in ISO-8859. The transcription does not include audio-text alignment information.

\begin{table}[h]
\caption{VOTE400 Dialog Speech Audio Format}
\label{dialog_speech_audio_format}
\begin{center}
\begin{tabular}{ll}
\hline
Property          & Value   \\ \hline \hline
Format.           & PCM               \\ \hline
Format Settings   & Little/Signed               \\ \hline
Codec ID          & 1                    \\ \hline
Bit Rate Mode     & Constant           \\ \hline
Bit Rate.         & 256               \\ \hline
Channel(s)        & 1               \\ \hline
Sampling Rate     & 16 kHz                   \\ \hline
Bit Depth         & 16 bits                      \\ \hline
\end{tabular}
\end{center}
\end{table}

The file name of each recording follows the pattern of \texttt{<P-ID>\_<G>\_<A>\_<R>\_<DT>}, where \texttt{P-ID} is a unique participant ID; \texttt{G} is a gender value (\texttt{F} for female, \texttt{M} for male), \texttt{A} is a age value, \texttt{R} is a regional code, and \texttt{DT} is the data-time of the recording session.

Participants and speech audio statistics for VOTE400 dialog dataset are shown in table \ref{vote400_dialog_stat_participants} and table \ref{vote400_dialog_stat_speech_audio}.

\begin{table}[h]
\caption{Demographics of VOTE400 Dialog Speech}
\label{vote400_dialog_stat_participants}
\begin{center}
\begin{tabular}{lcr}
\hline
Region(\texttt{R})     & No. Participants & Age ($\mu/\sigma$)  \\ \hline \hline
Seoul-si(SE) & 251(F:210,M:41) & 78.98/5.13        \\ \hline
Daegu-si(DG) & 108(F:95,M:13) & 80.33/6.08         \\ \hline
Gyoungki-do(GG) & 110(F:83,M:27) & 80.17/5.41      \\ \hline
Chungcheongnam-do(CN) & 6(F:6,M:0) & 77.00/3.69    \\ \hline
Jeollanam-do(JN) & 70(F:56,M:14) & 80.76/4.90      \\ \hline
Busan-si(PS) & 160(F:137,M:23) & 78.70/5.51        \\ \hline
Daejeon-si(DJ) & 96(F:72,M:24) & 78.81/5.24        \\ \hline
Gangwon-do(GW) & 109(F:94,M:15) & 80.07/5.50       \\ \hline
Gyeongsangbuk-do(GB) & 98(F:95,M:3) & 80.87/4.48   \\ \hline
Gwangju-si(GJ) & 87(F:70,M:17) & 79.39/5.77        \\ \hline
Chungcheongbuk-do(CB) & 17(F:17,M:0) & 80.47/5.51  \\ \hline
Ulsan-si(WS) & 58(F:49,M:9) & 76.97/4.48           \\ \hline  \hline
Total & 1,170(F:984,M:186) &  79.47/5.37           \\ \hline
\end{tabular}
\end{center}
\end{table}

\begin{table}[h]
\caption{Speech Audio Statistics for VOTE400 Dialog Speech}
\label{vote400_dialog_stat_speech_audio}
\begin{center}
\begin{tabular}{lrr}
\hline
Region(\texttt{R})     & Len.(secs) & Len.($\mu/\sigma$)  \\ \hline \hline
Seoul-si(SE) & 151,010 & 601.63/239.83               \\ \hline
Daegu-si(DG) & 60,740 & 562.42/228.14                \\ \hline
Gyoungki-do(GG) & 107,935 & 981.23/357.19            \\ \hline
Chungcheongnam-do(CN) & 5,193 & 865.62/293.98        \\ \hline
Jeollanam-do(JN) & 81,767 & 1,168.10/294.85           \\ \hline
Busan-si(PS) & 200,207 & 1,251.30/255.85              \\ \hline
Gangwon-do(GW) & 95,420 & 875.42/158.18              \\ \hline
Daejeon-si(DJ) & 123,138 & 1,282.70/293.83            \\ \hline
Gyeongsangbuk-do(GB) & 71,175 & 726.28/308.80        \\ \hline
Gwangju-si(GJ) & 92,699 & 1,065.52/276.53             \\ \hline
Chungcheongbuk-do(CB) & 20,135 & 1,184.41/309.54      \\ \hline
Ulsan-si(WS) & 70,754 & 1,219.90/254.43               \\ \hline \hline
Total & 1,080,179 &  923.23/380.17           \\ \hline
\end{tabular}
\end{center}
\end{table}

\subsection{Read Speech Data}
The total number of elderly participants is 104 and the total length of recordings is 100 hours. Table \ref{vote400_read_stat} shows the statistics of VOTE400 read speech data.

Audio format of the VOTE400 read speech data is as shown in \ref{read_speech_audio_format}, which is slightly different from the format of the dialog speech data.

\begin{table}[h]
\caption{VOTE400 Read Speech Audio Format}
\label{read_speech_audio_format}
\begin{center}
\begin{tabular}{ll}
\hline
Property          & Value   \\ \hline \hline
Format.           & PCM               \\ \hline
Format Settings   & Little/Signed               \\ \hline
Codec ID          & 1                    \\ \hline
Bit Rate Mode     & Constant           \\ \hline
Bit Rate.         & 705.6 kb/s               \\ \hline
Channel(s)        & 1               \\ \hline
Sampling Rate     & 44.1 kHz                   \\ \hline
Bit Depth         & 16 bits                      \\ \hline
\end{tabular}
\end{center}
\end{table}

\begin{table}[h]
\caption{Regional Distributions of VOTE400 Read Dataset}
\label{vote400_read_stat}
\begin{center}
\begin{tabular}{lrrr}
\hline
Region(\texttt{G}) & No. Persons & No. Sent. & Len.($\mu/\sigma$) \\ \hline \hline
Gyeongsangnam-do(GB) & 20 & 22,575 & 3.18/1.38       \\ \hline
Seoul-si(SE) & 18 & 19,220 & 3.31/1.49               \\ \hline
Jeollanam-do(JN) & 21 & 21,393 & 3.36/1.52           \\ \hline
Daegu-si(DG) & 25 & 26,950 & 3.60/1.87               \\ \hline
Gangwon-do(GW) & 20 & 21,676 & 2.73/1.12             \\ \hline \hline
Total & 104 & 111,814 & 3.25/1.54          \\ \hline
\end{tabular}
\end{center}
\end{table}

The file name of read speech data follows the pattern of \texttt{PID\_<P-ID>\_<DATE>\_<SENTENCE-NO>\_<R>}, where \texttt{P-ID} is a unique participant ID, \texttt{DATA} is the date of recording, \texttt{SENTENCE-NO} is a serial number put to each of the recorded sentences, and \texttt{R} is the region code as shown in table \ref{vote400_read_stat}. Each WAV file contains a single sentence, accompanied by a transcription text file encoded in \texttt{EUC-KR}.

Though the number of sentences chosen and presented to the participants was originally 2,250, the final total number of unique sentences in VOTE400 read speech data is 7,832, due to mistakes and slight variations in real utterances by older adults.

\section{Preliminary Experiment}
We conducted a preliminary experiment by training a MINDs Lab Inc.'s proprietary baseline speech recognizer(\texttt{M}), which is based on LSTM architecture, and estimating the STT accuracy using VOTE400. After fine-tuning the baseline with 50 hours each of dialog speech data and read speech data of VOTE400, a simple test with 100 sentences from different regions was performed and the results are as shown in table \ref{experiment}, along with the results when the sentences were tested on a commercial cloud-based STT engine(\texttt{C}).

\begin{table}[h]
\caption{STT Performance Test Results with VOTE400}
\label{experiment}
\begin{center}
\begin{tabular}{lrrr}
\hline
Region(\texttt{G}) & Gender & Acc. \texttt{M} (\%) & Acc. \texttt{C} (\%) \\ \hline \hline
SE & M & 90 & 90      \\ \hline
SE & F & 90 & 80      \\ \hline
GW & M & 80 & 90      \\ \hline
GW & F & 90 & 80      \\ \hline
DG & M & 70 & 80      \\ \hline
DG & F & 90 & 80      \\ \hline
GN & M & 90 & 80      \\ \hline
GN & F & 80 & 80      \\ \hline
JN & M & 70 & 50      \\ \hline
JN & F & 80 & 60      \\ \hline
\end{tabular}
\end{center}
\end{table}

\section{Summary}
We described a Korean speech dataset VOTE400 which is collected entirely from older adults of more than 75 years old. VOTE400 contains 300 hours of dialogue speech data and 100 hours of read speech data, with proficient varieties in gender and regions. To our knowledge, VOTE400 is by far one of the largest voice datasets that is oriented to voices of the elderly. We hope that this dataset will be useful to study older adult's voice features and realize voice technologies that work sufficiently well in elderly-care robotics.

\section*{ACKNOWLEDGMENT}
This  work  was  supported  by  the  Institute  of  Information    communi-cations Technology Planning   Evaluation(IITP) grant funded by the Koreagovernment(MSIT) (No. 2017-0-00162, Development of Human-care RobotTechnology for Aging Society)

\end{document}